\documentclass[12pt,a4paper]{article}
\usepackage{graphicx}

\def\qed{{\unskip\nobreak\hfil\penalty50\hskip .001pt \hbox{}\nobreak\hfil
          \vrule height 1.2ex width 1.1ex depth -.1ex
          \parfillskip=0pt\finalhyphendemerits=0\medbreak}\rm}

\def\Theorem #1. {\bigbreak\vskip-\parskip\noindent{\bf  Theorem   #1.}
    \quad\it}
\def\Corollary #1. {\bigbreak\vskip-\parskip\noindent{\bf Corollary #1.}
   \quad\it}
\def\Proposition #1. {\bigbreak\vskip-\parskip\noindent{\bf Proposition #1.}
   \quad\it}

\def\Proof#1.{\rm\par\ifdim\lastskip<\bigskipamount\removelastskip\fi
    \smallskip\noindent{\bf Proof.}\quad}
\def\R{{{\rm I} \! {\rm R}}}

\begin{document}
\vspace*{2cm}
\begin{center}
 \Huge\bf
Certain conclusions of Gordon decomposition
\vspace*{0.25in}

\large

Elena V. Palesheva
\vspace*{0.15in}

\normalsize

Omsk State University \\
644077 Omsk-77 RUSSIA
\\
\vspace*{0.5cm}
E-mail: palesheva@univer.omsk.su  \\
\vspace*{0.5cm}
\vspace{.5in}
ABSTRACT
\end{center}
We use a new view to the our reality which is presented by
Guts-Deutsch multiverse. In this article, we consider some conclusions of
Gordon decomposition of Dirac current.

\newpage

\setcounter{page}{1}


\section*{Introduction}

In paper \cite{1} we suppose that ghost spinors are same to Deutsch shadow
particles \cite{2}.
From this suggestion some questions are appeared. In this article such questions
are investigated. We assume that our reality is the Guts-Deutsch multiverse
\cite{2,3}. In paper \cite{4} some aspects of quantum particles interference were
studied. In result the study of shadow particle current is interested.
The last exists because in \cite{2} we have a new explanation of interference
in known quantum mechanics experiments. Deutsch assumes that shadow electrons
act upon a real electron. Herewith shadow electrons are electrons in
parallel universes and they are not observed by detectors. The mathematical
model of Guts-Deutsch multiverse was presented in \cite{3}. So
a problem of studied of shadow particles, or in other words ghost spinors,
is appeared.

\section{Gordon decomposition}

Let us consider Dirac equation
$$
i\hbar {\gamma}^k\left(\frac{\partial\psi}{\partial x^{\scriptscriptstyle k}}-{\Gamma}_k\psi
\right)-mc\psi =0,
$$
the spin connection ${\Gamma}_k$ is founded by formula:
$$
{\Gamma}_k=\frac{1}{4}g_{ml}\left(\frac{\partial{\lambda}^{(s)}_r}{\partial x^{\scriptscriptstyle
k}}\,{\lambda}^l_{(s)}-{\Gamma}^l_{rk}\right)s^{mr},
$$
where
$$
s^{mr}=\frac{1}{2}\left({\gamma}^m{\gamma}^r-{\gamma}^r{\gamma}^m\right).
$$
Moreover we take into account that
\begin{equation}\label{0}
{\gamma}^k\equiv{\lambda}^k_{(i)}{\gamma}^{(i)},
\end{equation}
here ${\lambda}^k_{(i)}$ is a i-th vector of tetrad and ${\gamma}^{(i)}$ are
Dirac matrixes for which we have the next  presentation with matrixes of
Pauly:
$$
{\gamma}^{(0)}=\left[\begin{array}{cc}I&0\\ 0&-I\end{array}\right],\quad{\gamma}^{(\alpha)}=
\left[\begin{array}{cc}0&{\sigma}_{\alpha}\\
-{\sigma}_{\alpha}&0\end{array}\right],
$$
$$
{\sigma}_1=\left[\begin{array}{cc}0&1\\ 1&0\end{array}\right],
{\sigma}_2=\left[\begin{array}{cc}0&-i\\ i&0\end{array}\right],
{\sigma}_3=\left[\begin{array}{cc}1&0\\ 0&-1\end{array}\right],
I=\left[\begin{array}{cc}1&0\\ 0&1\end{array}\right].
$$
Herewith the stress-energy tensor is defined by expression:
$$
T_{ik}=\frac{i\hbar c}{4}\left\{{\psi}^*{\gamma}^{(0)}{\gamma}_i\left(\frac{\partial\psi}
{\partial
x^{\scriptscriptstyle k}}-{\Gamma}_k\psi\right)-\left(\frac{\partial {\psi}^*}{\partial x^{
\scriptscriptstyle
k}}{\gamma}^{(0)}+{\psi}^*{\gamma}^{(0)}{\Gamma}_k\right){\gamma}_i\psi+\right.
$$
\begin{equation}\label{1}
\left.+{\psi}
^*{\gamma}^{(0)}{\gamma}_k\left(\frac{\partial\psi}{\partial x^{\scriptscriptstyle i}}
-{\Gamma}_i
\psi\right)-\left(\frac{\partial {\psi}^*}{\partial x^{\scriptscriptstyle i}}{\gamma}^{(0)}+{
\psi}^*{\gamma}^{(0)}{\Gamma}_i\right){\gamma}_k\psi\right\}.
\end{equation}

For two solutions of Dirac equation ${\psi}_1(x)$ and ${\psi}_2(x)$ we have
Gordon decomposition \cite[p. 45]{5}:
$$
c{\psi}_2^*{\gamma}^{(0)}{\gamma}^{(k)}{\psi}_1=\frac{i\hbar}{2m}\left[
{\psi}_2^*{\gamma}^{(0)}\frac{\partial{\psi}_1}{\partial x^k}-
\frac{\partial{\psi}_2^*}{\partial x^k}{\gamma}^{(0)}{\psi}_1\right]+
\frac{\hbar}{2m}g_{lm}\frac{\partial}{\partial
x^m}\left[{\psi}_2^*{\gamma}^{(0)}{\sigma}^{kl}{\psi}_1\right],
$$
where ${\sigma}_{ik}=\frac{i}{2}[{\gamma}_i,{\gamma}_k].$ As known the Dirac
current is defined by expression:
$$
j^{(k)}=c{\lambda}^{(k)}_i{\psi}^*{\gamma}^{(0)}{\gamma}^i{\psi}.
$$
Let us use the formula (\ref{0}), then first expression for Dirac current
will take a form:
$$
j^{(k)}=c{\lambda}^{(k)}_i{\psi}^*{\gamma}^{(0)}{\lambda}^i_{(m)}{\gamma}^{(m)}
{\psi}.
$$
If we take into consideration the property of tetrad vectors
${\lambda}^{(k)}_i{\lambda}^i_{(m)}={\delta}^k_m$ and summarize last
expression then we have
$$
j^{(k)}=c{\psi}^*{\gamma}^{(0)}{\gamma}^{(k)}{\psi}.
$$

Now let us take ${\psi}_1={\psi}_2$ then we will get the
Gordon decomposition of
Dirac current \cite{5}
\begin{equation}\label{2}
j^{(k)}=\frac{i\hbar}{2m}\left[
{\psi}^*{\gamma}^{(0)}\frac{\partial\psi}{\partial x^k}-
\frac{\partial{\psi}^*}{\partial x^k}{\gamma}^{(0)}\psi\right]+
\frac{\hbar}{2m}g_{lm}\frac{\partial}{\partial
x^m}\left[{\psi}^*{\gamma}^{(0)}{\sigma}^{kl}\psi\right].
\end{equation}
The first summand is a relativistic analog of displacement
current
$$
j=\frac{i\hbar}{2m}\left[\varphi\nabla{\varphi}^*-{\varphi}^*\nabla\varphi\right],
$$
here $\varphi$ is a solution of Shr\"{o}dinger equation. The second summand
in (\ref{2}) corresponds to spin current \cite{5}. So we have following.
A Dirac current and a momentum are not proportionals  to one another.
We will show that in some cases the displacement
current of ghost spinors equals to zero though the Dirac current of ghost
spinors not vanishs.

\section{Shadow particles currents}

Shadow particles are solutions of Dirac equation. Ones are defined by
vanished stress-energy tensor and a non-zero Dirac current. Now if we take
a track of the stress-energy tensor (\ref{1}) and also if we take into account
the Dirac equation, then the next result will be take placed \cite{6}:
\begin{equation}\label{4}
T^i_i=mc^2{\psi}^*{\gamma}^{(0)}\psi .
\end{equation}
Herewith we used the conjugated Dirac equation:
$$
i\hbar\left(\frac{\partial {\psi}^+}{\partial x^{\scriptscriptstyle
k}}+{\psi}^+{\Gamma}_k\right){\gamma}^k=-mc{\psi}^+,
$$
where we taken the standard denotation of Dirac-conjugated spinor
${\psi}^+={\psi}^*{\gamma}^{(0)}.$ In case of solutions of Dirac equation
for shadow particles the stress-energy tensor equals to zero. Then the track of the
last is equal to zero too. So from formula (\ref{4}) the next result was got.

\Proposition 1. \begin{bf}(Ghost spinor necessities condition)\end{bf}\\
If a solution of Dirac equation be
a ghost spinor then
$$
{\psi}^+\psi={\psi}^*{\gamma}^{(0)}\psi=0.
$$
\qed

Hereinafter we will take into consideration Minkowsky spacetime. So we have
that
${\gamma}^k={\gamma}^{(k)}$ and the spin connection ${\Gamma}_k$ is equal to
zero.

The next theorem was proved in \cite{4}.
\Theorem 1.
Let $\psi=u\cdot G(x)$ be a solution of Dirac equation. Herewith a spacetime
geometry is defined by Minkowsky metric. Let ${\psi}^*\psi\neq 0$ and
$$
G(x)=f(x) +i\cdot g(x),
$$
where $f(x)$ and $g(x)$ are smooth real functions. Moreover
$$
u=\left[\begin{array}{l} u_0\\ u_1\\ u_2\\
u_3\end{array}\right],
$$
where $\forall\quad i\quad u_i\in \R .$ In considered conditions $\psi$ will be a
ghost spinor if and only if $g(x)=a\cdot f(x),$ where $a=const\in \R .$
\qed

So if a solution of Dirac equation is presented by expression $\psi=u\cdot G(x)$ then it
will be a ghost spinor if and only if
 $G(x)$ is a real function. For these wave functions
the next statement about current, which is defined by expression
\begin{equation}\label{3}
j^{\,k}_p=\frac{i\hbar}{2m}\left[
{\psi}^*{\gamma}^{(0)}\frac{\partial\psi}{\partial x^k}-
\frac{\partial{\psi}^*}{\partial x^k}{\gamma}^{(0)}\psi\right],
\end{equation}
can be formulated.

\Theorem 2. Let a solution of Dirac equation $\psi$ be a ghost spinor.
Also let us assume that $\psi$ satisfies one of the next conditions:\\
\hspace*{0.5cm} 1) $\psi =u\cdot f(x),$ where $x=(x^0,x^1,x^2,x^3),$ herewith
$f(x)$ is a real function and bispinor $u$ has complex components,\\
\hspace*{0.5cm} 2) $\forall\, k$ $\partial\psi /\partial x^k=g_k(x)\psi,$
where $g_k(x)$ are complex functions,\\
then the displacement current in Gordon decomposition of Dirac current is equal to zero
everywhere.
\Proof.
Let us take into account the formula (\ref{3}). It is not difficult to see
that we must prove only the equality
\begin{equation}\label{5}
{\psi}^*{\gamma}^{(0)}\frac{\partial\psi}{\partial x^k}-
\frac{\partial{\psi}^*}{\partial x^k}{\gamma}^{(0)}\psi =0.
\end{equation}

Let a first case will be take placed. Then a shadow particle will be defined by
expression $\psi =u\cdot f(x).$ Thus if we insert corresponded solution of
Dirac equation to (\ref{5}) then we get the expression
$$
u^*{\gamma}^{(0)}u\left(f\frac{\partial f}{\partial x^k}-\frac{\partial
f}{\partial x^k}f\right)=0.
$$
So in this case the displacement current is equal to zero.

Let now a second case will be take placed:
$$
\forall\, k\quad\frac{\partial\psi}{\partial x^k}=g_k(x)\psi.
$$
Then the left part of the equality (\ref{5}) has the form
\footnote{We have indication $\overline{g}(x)$ for complex conjugation
function}:
$$
g_k(x){\psi}^+\psi-\overline{g_k}(x){\psi}^+\psi .
$$
In general case, for example if we take real particle, this expression
not equals to zero. For shadow particles, as we have proposition 1, both
summands equal to zero. The proof is finished.
\qed

From these results some corollaries are appeared. Let us consider two solutions
of Dirac equation ${\psi}_1$ and ${\psi}_2.$ If these particles interact
then for current of the result wave $\psi={\psi}_1+{\psi}_2.$
Both Dirac current and displacement current are disintegrated to the sum which contains
summands of free particles and summand of interaction. If for example
${\psi}_2$ is a ghost spinor then the next result exists.

Let a displacement current $j^k_{ip}$ corresponds to the wave
function ${\psi}_i.$ So the displacement current of the result wave satisfies to
$$
j^k_p=j^k_{1p}+j^k_{2p}+j^k_{12p},
$$
where in our case $j^k_{2p}=0$ and third summand is defined by interaction of
fields. Herewith
\begin{equation}\label{6}
j^k_{12p}=\frac{i\hbar}{2m}\left[
{{\psi}_1}^*{\gamma}^{(0)}\frac{\partial{\psi}_2}{\partial x^k}-
\frac{\partial{{\psi}_1}^*}{\partial x^k}{\gamma}^{(0)}{\psi}_2+
{{\psi}_2}^*{\gamma}^{(0)}\frac{\partial{\psi}_1}{\partial x^k}-
\frac{\partial{{\psi}_2}^*}{\partial x^k}{\gamma}^{(0)}{\psi}_1\right]
\end{equation}
and in general case the corresponding expression not equals to zero.
Thus we have $j^k_p\neq j^k_{1p}.$ So we can see that in considering case
interaction occurs at the expense of spin current.

\section{Wave-corpuscle duality}

As known, quantum mechanics is based on corpuscle and wave properties of
quantum particles. Let us assume that our reality is a multiverse. Then
corresponded duality may be explained by following.

In first let us notice that a square of module of probability amplitude of
shadow particle, which is a {\it own} particle of the some {\it real}
particle, is
equal to a square of module of probability amplitude of the corresponded
real particle.

Let "wave" consists of {\it one} real particle and {\it many}
own shadow particles. A shadow particle not can be detected. So in any moment
of time only one particle can be detected, it be a real particle. This will
correspond to corpuscle properties  of our "wave." Herewith such wave
property as interference will be explained by shadow paticles existance
\cite{1,2,3,4}.

\section*{Conclusion}

If a shadow particle interacts with a real particle then the displacement
current of a real particle is changed, but the displacement current of a
shadow
particle, in considering cases, is equal to zero. The displacement current
is in proportion to the momentum. So we can conclude that the interaction,
which is between a real particle and a shadow particle,
acts upon a location of a real particle in space. In spite of the fact that
the displacement current of a shadow particle
and the momentum of a shadow particle are not equal
to zero. But as for real as for shadow particle the Dirac current,
which is conservable value, be non-zero.
This influence will be exist as the spin current of a shadow particle is
not equal to zero. These results are
 agreed with Deutsch assumptions \cite{2}.

In paper \cite{1} in Minkowsky spacetime, ghost spinors were found. In \cite{7,8,9,10}
in curved spacetimes, ghost neutrinos were found. So we can see that
not only in flat spacetime the shadow particles can exist.


\end{document}